\documentstyle[a4,12pt]{article}

\begin{document}

\thispagestyle{empty}

\begin{flushright}
\bf{FIAN/TD-25/96}
\end{flushright}

\begin{center}
{\large {\bf {Factorial and Cumulant Moments in a Simple Cascade Model} }}
\end{center}

\medskip \medskip

\begin{center}
A.V.~Leonidov and D.M.~Ostrovsky
\end{center}

\bigskip

\begin{center}
{\bf Abstract}

Factorial and cumulant moments in a simple cascade models are considered.
Their characteristic features are shown to be similar to those observed in
QCD jets.
\end{center}

\newpage

\section{Introduction}

The characteristics of multiplicity distributions in multiparticle dynamics
has recently gained much attention due to a QCD analysis of generating
functional of these distributions and their moments. A very clear review of
this subject can be found in \cite{D1}, where one can also find the relevant
references, see also a recent review \cite{D2}. The main focus in the above
studies was initially a striking QCD prediction on the oscillatory behavior
of cumulants as functions of their rank. Actually it turned out to be more
convenient to analyze a ratio of cumulants to factorial moments. The
experimental data from many reactions (see \cite{D2}) confirmed this
prediction. The situation seemed to be all the more interesting, that
conventional multiplicity distributions such as negative binomial one, did
not predict such oscillations. However, a subsequent analysis \cite{CDNT}
has shown, that other more phenomenological models such as dual parton model
\cite{DPM} or quark-gluon string model \cite{KGSM} are also explaining the
data. This brings us to a natural question. What is the real (basic) origin
of the oscillations of the cumulants? To answer this question we construct a
simple probabilistic cascade and look at the properties of the cumulants. We
find exactly the same oscillation pattern as predicted theoretically and
observed experimentally, thus confirming one of the conclusions of \cite
{CDNT}, that the basic origin of these oscillations is a cascading origin of
the underlying process of particle formation.

\section{The model.}

Let us consider a model of cascade with discrete steps in time. Initially
(at time $t=0$) we have only one ''particle''. When time changes from $t$ to
$t+1$ any particle existing at a time $t$ splits with probability $p$ (a
parameter of the model) into 2 particles or does not split with probability $%
1-p$.

Now $P_{n}(t)$ is a probability that on $t$ step $n$ particles were produced.
The ''first split'' equation reads
\begin{equation}
P_{n}(t)=(1-p)P_{n}(t-1)+p\sum%
\limits_{n_{1}=0}^{n}P_{n_{1}}(t-1)P_{n-n_{1}}(t-1)\quad ,\quad
P_{n}(0)=\delta _{n1}  \label{1split}
\end{equation}
Let us now consider a generating function related to the probabilities $%
P_{n}(t)$:
\begin{equation}
G(t,z)=\sum\limits_{n=0}^{\infty }P_{n}(t)(1+z)^{n}
\end{equation}
From  (\ref{1split}) we have
\begin{equation}
G(t,z)=(1-p)G(t-1,z)+pG^{2}(t-1,z)  \label{G(t,z)}
\end{equation}
The boundary conditions for this relation is\medskip \newline
$1.\ G(t,0)=1$ -- the total probability conservation;\newline
$2.\ G(0,z)=1+z$ -- corresponding to $P_{n}(0)=\delta _{n1}$

\section{Mean multiplicity and moments.}

By definition a mean multiplicity $n(t)$ is related to a generating function by
\begin{equation}
n(t)=\sum\limits_{n=0}^{\infty }nP_{n}(t)=\left. \frac{\partial G(t,z)}{%
\partial z}\right| _{z=0}.  \label{n(t)_intro}
\end{equation}
It gives for (\ref{G(t,z)})
\begin{equation}
n(t)=(1-p)n(t-1)+2pn(t-1)=(1+p)n(t-1)=(1+p)^{t}.  \label{n(t)}
\end{equation}
Last equality follows from $G(t,0)=1$ (see above) and $n(0)=1$.

Factorial moments are introduced as coefficients of $F_{q}(t)$ in the
expansion
\begin{equation}
G(t,z)=1+\sum\limits_{q=1}^{\infty }\frac{z^{q}}{q!}F_{q}(t)n^{q}(t).
\label{F(t)}
\end{equation}
After substitution into (\ref{G(t,z)}) we find
\begin{equation}
(1+p)^{q}F_{q}(t)=(1-p)F_{q}(t-1)+p\sum%
\limits_{k=0}^{q}C_{q}^{k}F_{k}(t-1)F_{q-k}(t-1)\,,\,q>1  \label{F(t)eq}
\end{equation}

It is easy to demonstrate that the limit $\lim\limits_{t\rightarrow \infty
}F_{q}(t)=F_{q}$ exists. Therefore from  (\ref{F(t)eq}) we get
\begin{equation}
F_{q}=\frac{1}{(1+p)^{q-1}-1}\frac{p}{1+p}\sum%
\limits_{k=1}^{q-1}C_{q}^{k}F_{k}F_{q-k}\,,\,q>1\qquad F_{1}=1  \label{Feq}
\end{equation}

Now it is possible to treat asymptotic factorial moments as factorial
moments of ''asymptotic'' generating function.
\begin{equation}
G(a)=1+\sum\limits_{q=1}^{\infty }\frac{F_{q}}{q!}a^{q}.  \label{Gass}
\end{equation}
The equation on $G(a)$ now reads
\begin{equation}
G((1+p)a)=(1-p)G(a)+pG^{2}(a)  \label{Gasseq}
\end{equation}
with the initial conditions $G(0)=1$ and $G^{\prime }(0)=1$.

\section{Cumulants and $H_{q}$}

Cumulants are introduced as coefficients $K_{q}(t)$ in the expansion
\begin{equation}
\ln G(t,z)=\sum\limits_{q=1}^{\infty }\frac{z^{q}}{q!}K_{q}(t)n^{q}(t).
\label{K(t)}
\end{equation}
The cumulant and factorial moments are actually not independent. The
relation between the two can be traced from the formula $%
G_{z}(\ln G)_{z}^{\prime }=G_{z}^{\prime }$. From (\ref{F(t)}) and (\ref{K(t)}%
) we obtain
\begin{equation}
K_{q}=F_{q}-\sum\limits_{k=1}^{q-1}C_{q-1}^{k}K_{k}F_{q-k}  \label{relation}
\end{equation}

It turns out convenient to consider the ratios $H_{q}=K_{q}/F_{q}$ obtained from
(\ref{Feq}) and (\ref{relation}). The resulting $H_q$  for some values of $p$
are shown
in Fig.~1. We see, that the ratio of cumulant to factorial moments $H_{q}$
in our simple cascade model provides the same oscillation pattern that
follows from a number of theoretical models and is observed experimentally.
How good is a quantitative predictions of the model when compared
to the experimental data is questionable. We restrict ourselves to
illustrate this situation at  Fig.~2.

It is instructive to look more attentively at two extreme limits of our
cascade model.\newline
\medskip
1. $p\rightarrow 0$\newline
Keeping the leading power of $p$ in (\ref{Gasseq}) we get
\begin{equation}
paG^{\prime }(a)=pG(a)(G(a)-1)  \label{p20eq}
\end{equation}
The solution of this equation that satisfies boundary conditions is
\begin{equation}
G(a)=\frac{1}{1-a}  \label{p20sol}
\end{equation}
so $F_{q}^{(0)}=q!$, $K_{q}^{(0)}=(q-1)!$ and $H_{q}^{(0)}=1/q$. It is worth
noting that this result is valid only for $pq\ll 1$ (not for $p\ll 1$).%
\newline
\medskip
\noindent 2. $p=1$\newline
Although in the context of our study this limit looks quite peculiar, let us
proceed and perform an explicit calculation, which gives
\begin{equation}
G(2a)=G^{2}(a).  \label{p21eq}
\end{equation}
The obvious solution of this equation is
\begin{equation}
G(a)=e^{a}  \label{p21sol}
\end{equation}
so $F_{q}^{(1)}=1$, $K_{q}^{(1)}=\delta _{q1}$ and $H_{q}^{(1)}=\delta _{q1}$%
. Such moments are produced by Poisson distribution.

\section{\protect\medskip The continuous limit.}

In the context of our model investigations it is useful to determine
continuous limit of (\ref{G(t,z)}). This allows to calculate not only
asymptotic shape of generating function and moments but also their
dependence on mean multiplicity $n$.

It is convenient to give (\ref{G(t,z)}) the form

\begin{equation}
G(t,z)-G(t-1,z)=pG(t-1,z)(G(t-1,z)-1).
\end{equation}
If $p$ is considered to be a small one can replace the difference in l.h.s.
of this equation by derivative (the change at one step becomes infinitesimal)

\begin{equation}
\frac{\partial G(t,z)}{\partial t}=pG(t,z)(G(t,z)-1)  \label{dG/dz}
\end{equation}
(we have also changed $t-1$ to $t$ ). In this limit the equation on mean
multiplicity is (see \ref{n(t)_intro})

\begin{equation}
\frac{dn(t)}{dt}=pn(t)
\end{equation}
that is
\begin{equation}
n(t)=e^{pt}n(0)=e^{pt}  \label{n(t)_lim}
\end{equation}
according to the initial condition $P_{n}(0)=\delta _{n1}$. Solution of (\ref
{dG/dz}) is clear
\begin{equation}
G(t,z)=\left(1-\frac{n(t)z}{1+z}\right)^{-1},  \label{G(t,z)_lim}
\end{equation}
where initial conditions and (\ref{n(t)_lim}) have been used. From this we
obtain factorial moments and cumulants
\begin{equation}
F_{q}(t)=q!\left( 1-\frac{1}{n(t)}\right) ^{q-1}
\end{equation}
\begin{equation}
K_{q}(t)=(q-1)!\left( \left( 1-\frac{1}{n(t)}\right) ^{q}-\frac{(-1)^{q}}{%
n^{q}(t)}\right)
\end{equation}
so
\begin{equation}
H_{q}(t)=\frac{1}{q}\left( 1-\frac{1}{n(t)}\right) \left( 1-\frac{(-1)^{q}}{%
(n(t)-1)^{q}}\right) .  \label{H(t)_lim}
\end{equation}
In the limit $n\rightarrow \infty $ this result becomes the same as in (\ref
{p20sol}).

 Here it is appropriate to discuss some relations between our model (in
continuous limit) and a PB (pure birth) one considered in \cite{Bij}. That
model is formulated as an equation on probability to obtain $n$ particles at
the moment of (continuous!) time $t$
\begin{equation}
\frac{\partial P(n,t)}{\partial t}=\lambda (n-1)P(n-1,t)-\lambda nP(n,t).
\label{Bij_P}
\end{equation}
For this equation on $P$'s the equation on generating function (which we
denote by $\tilde{G}$) is
\begin{equation}
\frac{\partial \tilde{G}}{\partial t}=\lambda z(1+z)\frac{\partial \tilde{G}%
}{\partial z}  \label{tild_G_eq}
\end{equation}
Mean multiplicity $\tilde{n}$ in this case satisfies
\begin{equation}
\frac{\partial \tilde{n}(t)}{\partial t}=\lambda \tilde{n}(t)=e^{\lambda t}%
\tilde{n}(0).  \label{tild_n}
\end{equation}
It is easy to see that the equation (\ref{tild_G_eq}) has a solution
\begin{equation}
\tilde{G}(t,z)=\tilde{G}\left( \frac{e^{\lambda t}z}{1+z}\right) =f\ \left(
\frac{\tilde{n}(t)z}{1+z}\right) .  \label{tild_G}
\end{equation}
Here $f$ is a function determined by initial ($t=0$) distribution. Universal
conditions on $f$ are only $f(0)=1$ (from $\tilde{G}(t,z=0)=1$) and $%
f^{\prime }(0)=1$ (from $\partial \tilde{G}(t,z=0)/\partial z=\tilde{%
n}(t)$). So in this model the asymptotic ($\tilde{n}\rightarrow \infty $)
behavior of factorial moments and cumulants is determined by initial
distribution while the dependence on $\tilde{n}$ is described by differential
equations (\ref{Bij_P}) and (\ref{tild_G_eq}).

It is clear that in the continuous limit our model turns into PB if one
specifies some initial conditions to (\ref{G(t,z)}).

\section{Conclusions}

From the study of the simple cascade model of particle multiplication we
see, that it correctly reproduces the oscillation pattern predicted by a
number of theoretical schemes and observed experimentally. We conclude that
the basic origin of such oscillations is a cascade structure of underlying
dynamic picture.  This cascade could in principle originate both from the
quark-gluon phase of the evolution of hadronic system and from the
nonperturbative hadronization stage.

We are grateful to I.M.~Dremin and I.V.~Andreev for useful discussions
and suggestions.
This study was supported by Russian Foundation for Basic Research under
Grant 96-02-16210a.


\begin{thebibliography}{9}
\bibitem{D1}  I.M.~Dremin, Usp.\ Fiz.\ Nauk\ 164 (1994), 785

\bibitem{D2}  I.M.~Dremin, Multiplicity Moments in QCD and Experiment,
hep-ph/9604245, 1996

\bibitem{CDNT}  A.~Capella, I.M.~Dremin, V.A.~Nechitailo and J.~Tran Thanh
Van, Moment Analysis of Multiplicity Distributions, hep-ph/9604247, 1996,
 {\it{Zeit.\ Phys.}}\ {\bf{C}}, to be published

\bibitem{DPM}  A.~Capella, J.~Tran Thanh Van, Z.\ Phys.\ C23 (1984), 165;
A.~Capella et al., Phys.\ Rev.\ D35 (1987), 2921

\bibitem{KGSM}  A.B.~Kaidalov, Sov.\ J.\ Nucl.\ Phys.\ 45 (1987), 902

\bibitem{Bij} M.~ Biyajima et al. Phys. Rev. Lett. B237 (1990), 563
\end{thebibliography}
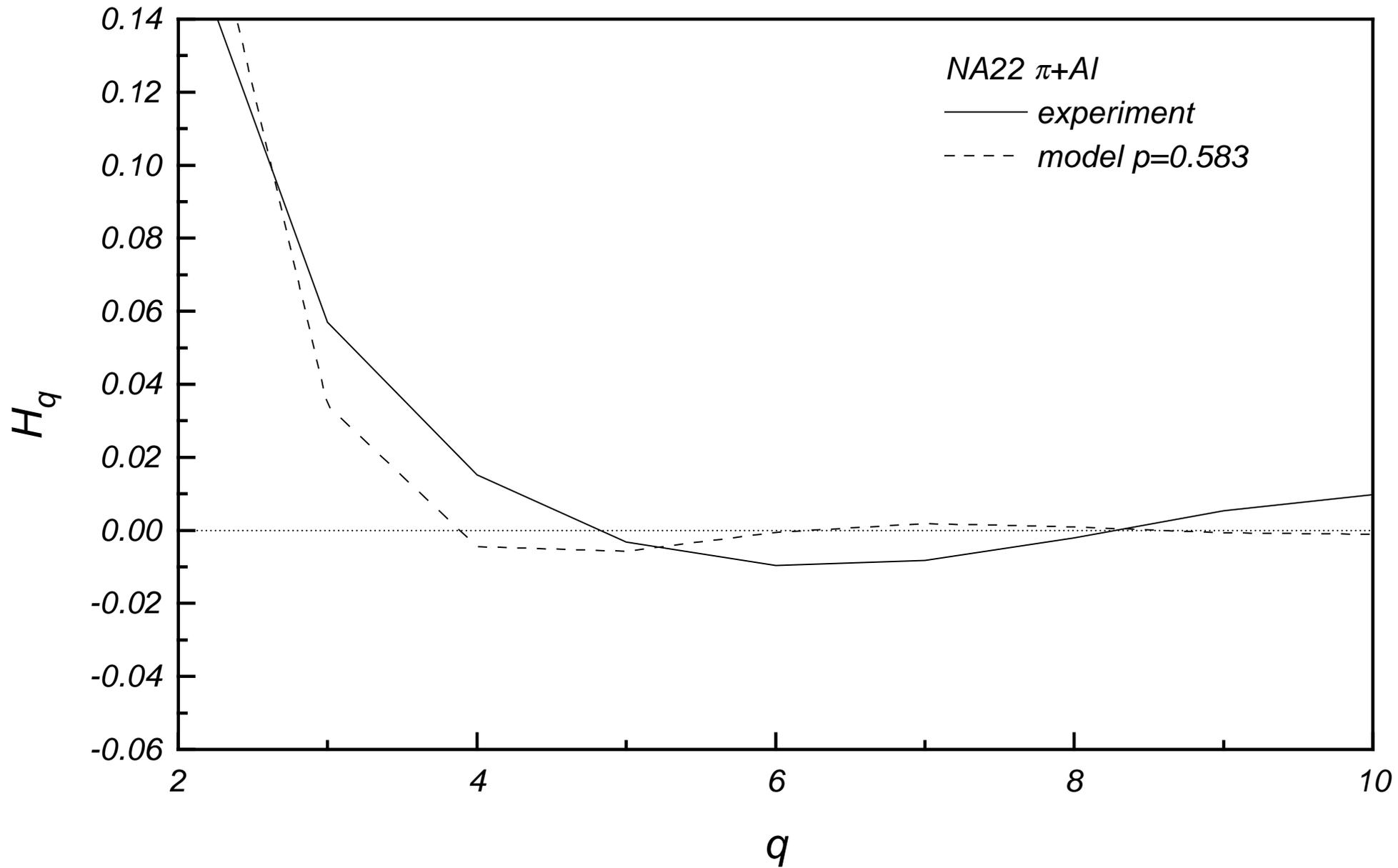
\end{document}